\documentclass{osa-article}

\journal{oe}
\articletype{Research Article}

\usepackage{lineno}
\usepackage{todonotes}

\begin{document}

\title{High-availability displacement sensing with multi-channel self mixing interferometry}

\author{Robin Matha \authormark{1,2}, Stéphane Barland \authormark{2,*} and François Gustave\authormark{1}}

\address{\authormark{1}DOTA, ONERA, Université Paris-Saclay, F-91123, Palaiseau, France\\
\authormark{2}Université Côte d'Azur - CNRS, Institut de Physique de Nice, 1361 route des Lucioles, F-06560, Valbonne, France}

\email{\authormark{*}stephane.barland@univ-cotedazur.fr} %

\newcommand{\steph}[2][noinline]{\todo[#1, color=blue!10, linecolor=blue!50!white, bordercolor=blue!50!white]{\small \texttt{Steph}: #2}}

\newcommand{\franc}[2][noinline]{\todo[#1, color=green!15, linecolor=green!50!white, bordercolor=green!50!white]{\small \texttt{Francois}: #2}}

\newcommand{\ffig}[1]{Fig.~\ref{#1}}

\begin{abstract}
	Laser self-mixing is in principle a simple and robust general purpose interferometric method, with the additional expressivity which results from nonlinearity. However, it is rather sensitive to unwanted changes in target reflectivity, which often hinders applications with non-cooperative targets. Here we analyze experimentally a multi-channel sensor based on three independent self-mixing signals processed by a small neural network. We show that it provides high-availability motion sensing, robust not only to measurement noise but also to complete loss of signal in some channels. As a form of hybrid sensing based on nonlinear photonics and neural networks, it also opens perspectives for fully multimodal complex photonics sensing.
\end{abstract}

\section{Introduction}

	Complex (nonlinear or disordered) systems have long been considered as high-capacity "information processors" \cite{nicolis2012foundations}. For the specific task of processing environmental changes (\textit{ie "sensing"}), complex waves and photonic systems are particularly useful thanks to their capability to process information from a distance. Recently, hybrid solutions leveraging complex wave physics and computer neural networks have enabled oustanding achievements \cite{caramazza2018neural,del2021deeply}, at the intersection between neural networks and optics \cite{mengu2022intersection}. Thanks to its nonlinearity, self-mixing interferometry can carry much more information than its linear counterpart, which makes it potentially useful for many sensing tasks  \cite{giuliani2002laser,kane2005unlocking,donati2012developing,taimre2016laser,li2017laser,rakic2019sensing,brambilla2020versatile}. 

Self mixing interferometry is based on a delayed feedback effect, where a laser beam re-enters the emitting laser itself (in most cases in a semiconductor laser) after being reflected on a target. As is well known, this feedback alters the operation point of the laser and by monitoring this operation point (either via an integrated photodiode or by simply measuring the voltage across the diode), information about the displacement of the target can be retrieved. Thanks to this simplicity and compacity, this scheme is expected to provide a robust and reliable sensing technique, for instance for the displacement of a target along the light propation axis. However, the operation range in which information can be reliably retrieved is in fact an important constraint, notably in terms of the amount of light which re-enters the emitting laser. This is often characterized via the coupling parameter $C$ which also depends on the target distance. We refer the reader to for instance \cite{giuliani2002laser} for a very complete discussion of the different feedback regimes depending on $C$ (including more parameters in \cite{bertling2022feedback}) but here we only underline that in general, the shape of the signal strongly depends on $C$, which makes signal processing difficult. More importantly, when the amount of light re-entering the device is too small ($C<<1$), the self mixing signal becomes harmonic and the information about the direction of the displacement of the target is lost. At the other edge of the operation regime (when $C>4.6$), the dynamics of the laser with reinjection becomes multistable or even unstable towards chaotic regimes. Again, in this range, the self-mixing signal does not carry reliable information about the displacement of the target. For these reasons, great care must be taken to keep the system within the useful operation range. Different approaches have been taken to mitigate unwanted variation of $C$ for instance due to speckle effect, including dedicated algorithms \cite{siddiqui2017all}, on the fly parameter estimation \cite{bernal2021toward,an2022measuring} or training a neural network in different alignment conditions \cite{barland2021convolutional,barland2022displacement}. However, the hardest limitation is the progressive loss (eventually up to complete lack) of significance of the signal when $C<<1$ or $C>4.6$. Since this limitation is of physical origin, tracking hardware has been proposed for the case of speckle \cite{bernal2014robust} and adapted beam focussing has been analysed for large displacements \cite{de2010laser}. 

Here, we propose that high-availability motion sensing can be achieved with a multi-channel self-mixing interferometer equipped with a simple, embeddable neural network. Multichannel self mixing has already been envisioned for complex measurements (see \textit{eg} \cite{ottonelli2009simultaneous,ottonelli2009laser,lim2009self,lim2010self}) but only seldomly considered as a potential enhancement \cite{tucker2014self} for the acquisition of a single measurement. On the other hand, machine learning has been increasingly used in self-mixing applications, including for fringe detection \cite{li2019deep,kou2020fringe,siddiqui2022fringe}, parameter estimation \cite{an2022measuring}, signal enhancement \cite{wei2007pre,ahmed2019self}, vibration measurement \cite{liang2022combined} and displacement inference \cite{barland2021convolutional}. 
Here we show that, thanks to the intrinsic capacity of neural networks to process high-dimensional data, a multichannel self-mixing sensor can provide high availability displacement measurements, robust against signal loss and with enhanced resolution.

In section \ref{sec:exp} we present the experimental arrangement (\ref{subsec:exp}), the neural network design (\ref{subsubsec:design}) and its training procedure (\ref{subsubsec:training}). We assess the performance of the system in section \ref{sec:results} in terms of accuracy (\ref{subsec:accuracy}), robustness against noise (\ref{subsec:robustness}) and measurement availability (\ref{subsec:availability}). We present our conclusions in section \ref{sec:conclusion}.

\section{Experimental set up, model and training}
\label{sec:exp}

\begin{figure}
	\centering
	\includegraphics[width=0.7\textwidth]{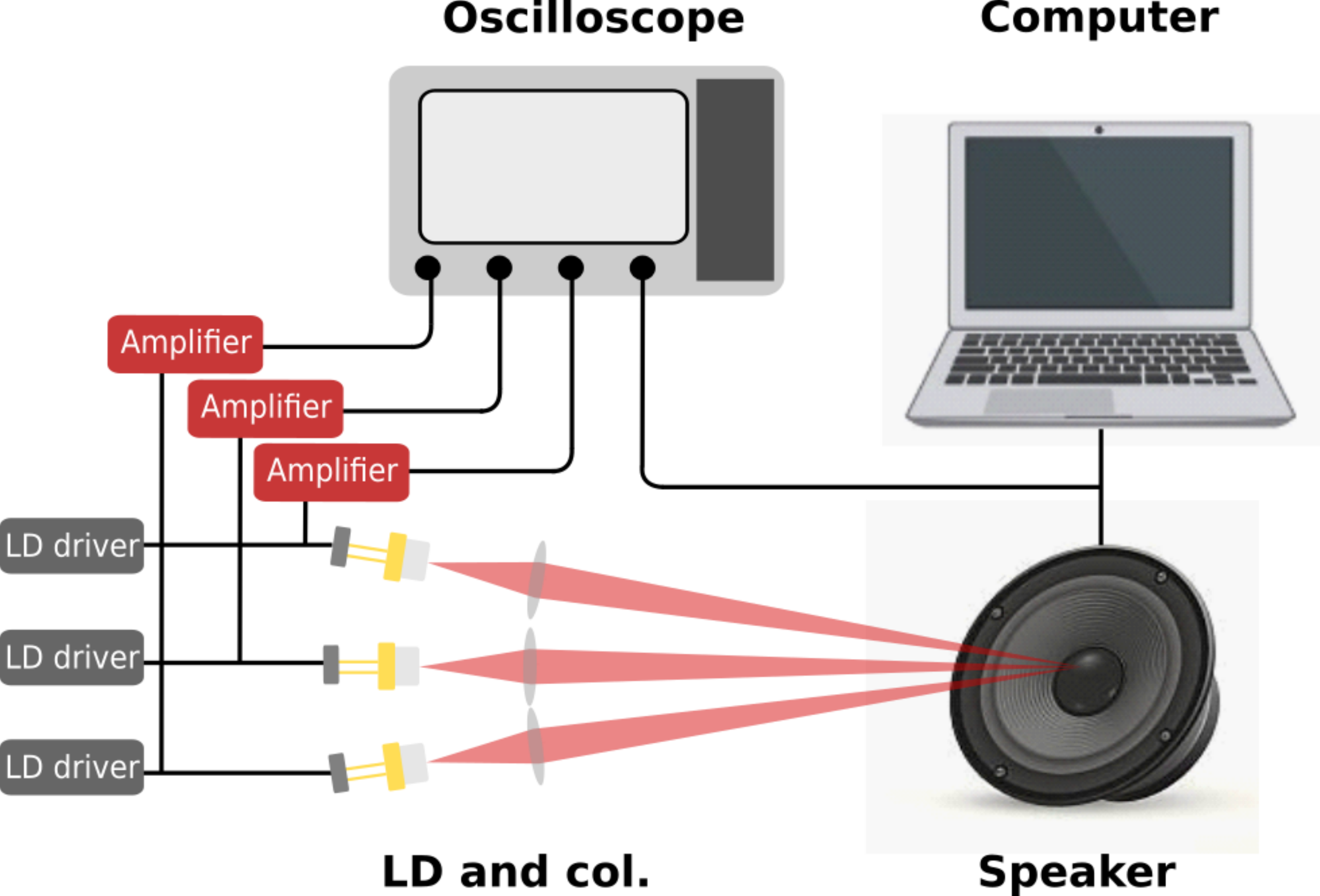}
	\caption{Scheme of the experimental setup. Three independent self-mixing channels monitor the displacement of a single non-cooperative target and a neural network processes the resulting high dimensional data to estimate the displacement of the target.
	\label{fig:setup}}
\end{figure}

\subsection{Experiment}
\label{subsec:exp}
The principle of the experimental arrangement is presented on \ffig{fig:setup}. It consists of three independent self-mixing measurement channels with a (calibrated) speaker which acts as a target, a moving surface. Each channel is composed of a power supply, a laser diode and a signal amplification stage. The lasers of channels 1 and 2 emit at $\lambda=1310$~nm (ML725B8F) an their coherent emission threshold is 6.5~mA. The laser of channel 3 emits at $\lambda$=1550~nm (ML925B45F) and its threshold is around 10~mA. For all the acquisitions that will be described later, the lasers were respectively driven with DC currents of 7.7mA (1 and 2) and 41.7mA (channel 3). The laser beams are focussed onto the speaker surface, which is located about 15cm away from each laser. The lasers are placed very close to each other to minimize the angle between the beams and the normal to the speaker surface, minimizing the error between the actual displacement and its projection along each beam propagation direction.

\begin{figure}
	\centering
	\includegraphics[width=0.7\textwidth]{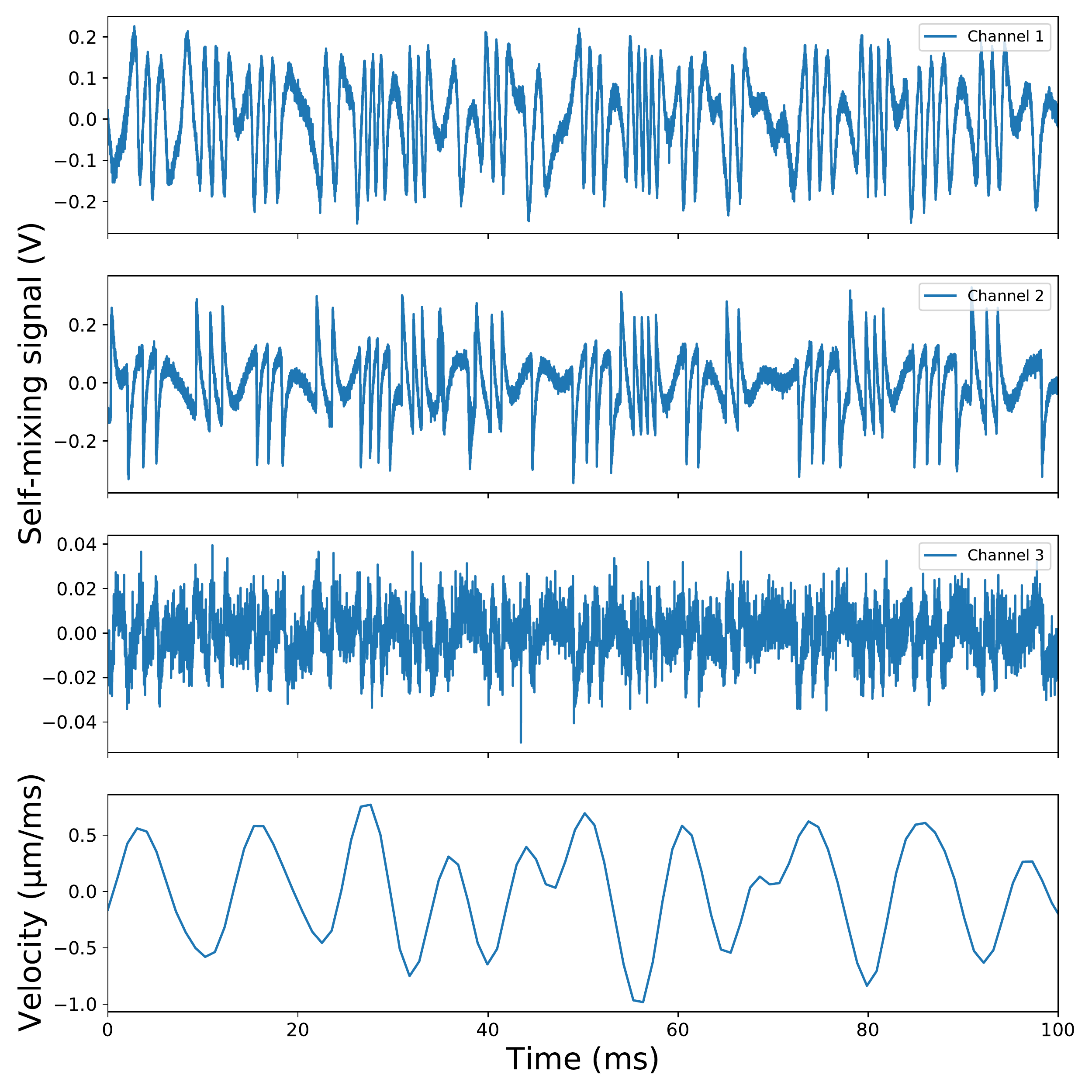}
	\caption{Experimental data example. The top three rows show the self-mixing signal provided by each laser and the bottom row the corresponding displacement. The lasers are purposefully set to provide different quality of self-mixing signals (see text).
	\label{fig:example}}
\end{figure}

The three self-mixing signals are obtained by measuring the voltage across each laser diode for two of them and the current through the internal photodiode for the third one (dictated by available equipment). These signals being of low amplitude, each voltage signal is amplified by an AC-coupled amplifier with $10^4$ gain factor and several MHz bandwidth and the photodiode current is amplified by a transimpedance amplifier. Since self-mixing interferometry with non-cooperative targets is often plagued with signals of varying quality (mostly due to speckle effect) we purposefully align the lasers slightly differently so that each laser operates in a slightly different regime. An example is shown on \ffig{fig:example}. On the top row, the self-mixing signal is not particularly noisy but it corresponds to a rather low reinjection value and therefore the asymmetry of the fringes is not very visible. On the second row, the signal is of excellent quality with low amount of noise and well defined asymmetric fringes. Finally, the signal shown on the third row is hardly exploitable at all, with a very poor signal to noise ratio.

The displacement of the speaker has been calibrated and the linearity of the response of the speaker to a harmonic signal in a range of 5 to 100~Hz allows us to have access to the effective displacement of the surface. In this frequency range the response of the loudspeaker does not show any phase shift. These harmonic signals are generated by the sound card of the computer with a normalized amplitude between 0.1 and 1, which means displacements between 3.5µm and 7.5µm. This electrical signal is recorded during the experiment and (after some preprocessing described below) forms the basis of the truth signal. An example of the displacement signal is shown on the bottom row of \ffig{fig:example}.

\subsection{Model and Training}
\subsubsection{Model design}
\label{subsubsec:design}
In this experiment, the neural network is supposed to infer the displacement of the target on the basis of three interferometric signals. The problem is therefore in principle a (multi-)sequence to sequence task. However, as was pointed at in \cite{barland2021convolutional}, defining a spectral band of operation for the displacements to be measured allows one to convert the task into a simpler regression task by downsampling the displacement signal. In this case, the frequency operation regime we define ranges from 10 to 100~Hz. Within this range of displacement frequency and with 4~$\mu$s sampling time for the signal acquisition (both interferometric signals and displacement), we choose to analyse the time series in chunks of 1.024~ms, each of them containing 256 data points per interferometric channel and per displacement measurement. Due to the selected frequency band and the Nyquist theorem, each chunk of 256 measurement of speaker voltage can be replaced by a single average displacement value over the 256 time steps. This converts the sequence to sequence task into a much simpler regression task, where one displacement value must be inferred from three interferometric sequences.
At this point, we can use the very simple convolutional neural network architecture proposed in \cite{barland2021convolutional}, since it appeared to carry sufficient information capacity to efficiently process single-channel data. However, each input vector is now composed of three channels, one per interferometric signal. Although advanced use of multiple sensing modalities is a very relevant area of research in itself (see \textit{eg}\cite{gadzicki2020early,khacef2020brain,dimitri2022short,liang2022foundations}), here we deal with the simpler case of multiple measurement channels which operate on similar modalities. Thus, we stick to a very simple "early fusion" approach: the first convolutional layer operates on the three input channels and fuses the resulting kernels into a single output space, irrespectively of what specific channel activated them. After this stage, the network consists of a contracting stack of downsampling and 1D, single channel, convolutional layers. At the end of the stack, two fully connected layers perform the final regression towards a single displacement value per time interval. All layers use a Rectified Linear Unit activation function except for the last layer which is linear. Details about the network are available in \nameref{sec:appendix}.

\subsubsection{Training}
\label{subsubsec:training}
The training set \cite{matha2023multichannel} consists only of experimental data. We record simultaneously the three self-mixing signals and the speaker voltage, calibrated as a displacement. The sampling rate is 4~$\mu$s and the record length is 499968 points. The training set features only harmonic displacements with eight evenly spaced frequencies between 53Hz and 93Hz and six evenly spaced amplitudes between 0.4 and 0.9~V applied to the speaker, corresponding to 3.5 and 7.5 $\mu$m. Thus, the training set contains 48 different configurations in terms of frequency and amplitude of periodic displacement. 

Before processing by the network, the interferometric signals are normalized and centered around zero by dividing them by their standard deviation and substracting the mean value of the signal over the full record length. In order to reinforce the robustness of the network to noise, we add to each signal a gaussian white noise of standard deviation $\sigma_n=\sigma_s$ where $\sigma_s$ is the standard deviation of the original self-mixing signal.

As for the truth signal, first we remove the electronic noise (intrinsic oscilloscope noise and digitalization noise) by smoothing the displacement signal using a Savitsky-Golay filter whose parameters are a sliding interval of 1001 points and a polynomial degree of 2. Then we compute the average displacement during a time window of 256 sampling points which gives a signal in units of Volts per time window. We then use the duration of the sampling rate $dt=4\mu$s which gives a time window of 1.024~ms and the speaker calibration of 31.3$\mu$m/V to convert the signal in physical units of $\mu m/ms$.

The network can then be trained using as input about $9.5*10^4$ arrays of size $3\times256$ corresponding to three interferometric signals over time windows of $256\times dt$ and as truth the corresponding average displacement over each time window, with 10\% of the samples kept for validation. Of course during training the order of samples is randomized at each epoch but an additional randomization is also operated on the order of the channels themselves, \textit{ie} to a single displacement value can correspond any permutation of the measurement channels. This point implies that the network must be trained in physical units of $\mu$m/ms but we have observed that it leads to a remarkable improvement of the reconstruction performance. We attribute it to an enhancement of training for all weights especially at the first layer instead of focusing training immediately on the most significant channel. In addition to randomization, during training we randomly replace one of the self-mixing signals with white noise (with probability 1/4) to prepare the network to potential channel loss during the inference phase.
We train the network by minimizing the mean squared error between the inferred and the true displacement during 18 epochs in batches of 32 samples. On a (consumer grade) GTX1080 GPU the training time is about 20~min.

\section{Results}
\label{sec:results}
\begin{figure}
	\centering
	\includegraphics[width=0.7\textwidth]{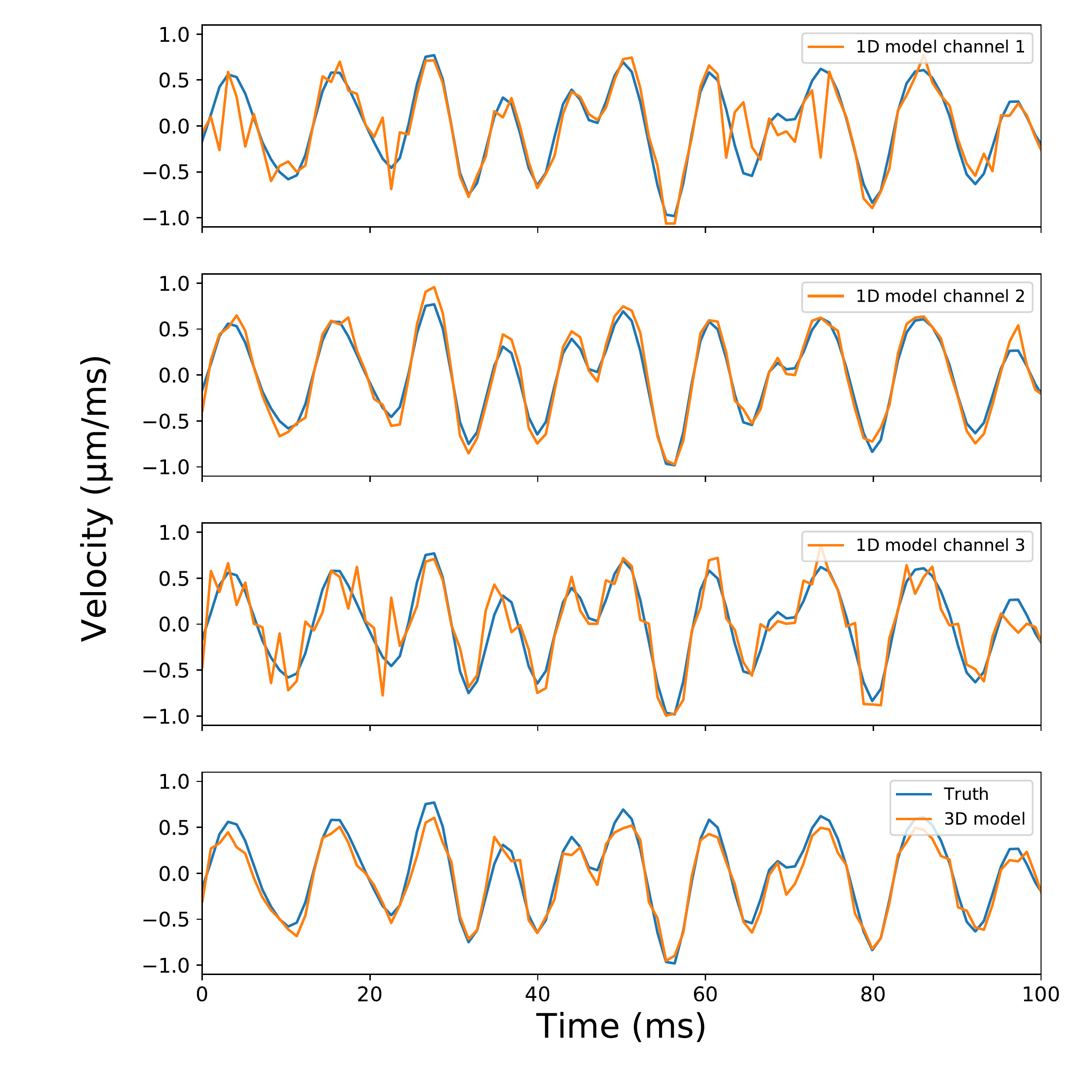}
	\caption{Reconstruction performance based on the single channels in isolation (the top three rows are three 1D models) and on the three channels simultaneously (bottom row, 3D model).
	\label{fig:models}}
\end{figure}

After training, we assess the performance of a multi-channel measurement approach against that of the usual single-channel approach, first in terms of accuracy and second in terms of robustness to measurement noise up to the complete loss of one or the other channel. Since each channel provides an independent measurement, we will refer to a model processing all three channels as a three-dimensional (3D) model, as opposed to a model processing a single dimension, which we will refer to as a one-dimensional (1D) model.

\subsection{Accuracy}
\label{subsec:accuracy}

On \ffig{fig:models}, we show the reconstruction of the target trajectory from each measurement channel and the reconstruction based on the three channels simultaneously. The displacement reconstruction is expected to operate correctly for arbitrarily complex displacements therefore we analyse its performance (as in \cite{barland2021convolutional}) on a random displacement with a prescribed bandwidth (here 10 to 100~Hz). To achieve this, we generate a random $\delta$-correlated signal which we Fourier filter with a fifth order Butterworth filter between 10 and 100~Hz. This signal is sent to the speaker, which then undergoes a random motion. On the top three traces, we use three 1D models trained on the experimental data set described above, each model processing only one of the three self-mixing signals. We quantify the quality of the reconstruction by measuring the Pearson correlation coefficient and the Root Mean Squared Error between the true displacement and the reconstruction on a 512-seconds long time trace of this random displacement ($5*10^5$ samples of 1.024~ms duration). On the top trace, the reconstruction is rather good with a correlation coefficient between the true displacement and the reconstruction of $r=0.83$ and a Root Mean Square Error $RMSE<0.23~\mu$m/ms. The second row is based on the best measurement channel, with low detection noise and well defined, clearly asymmetric fringes. Correspondingly, the quality of the target displacement reconstruction is excellent, with a correlation coefficient between the true displacement and the reconstruction of $r=0.97$ and a $RMSE<0.11~\mu$m/ms. On the third row instead, the reconstruction of the displacement is of again of lower quality, which is not unexpected since the self-mixing signal is really very poor (see \ffig{fig:example}, third row). We note however that the reconstruction is essentially equivalent to that of the first channel, with a correlation coefficient $r=0.83$ and $RMSE<0.24~\mu$m/ms. Finally, on the bottom row, we show the reconstruction based on a 3D model processing the three channels simultaneously. This reconstruction is almost as good as the one obtained only on the single highest quality measurement (channel 2) and is markedly better ($r=0.96, RMSE=0.13\mu$m/ms) than the one obtained on the two poorest channels (1 and 3).

\subsection{Robustness against signal degradation} 
\label{subsec:robustness}

We have seen above that (at least with the simple early fusion approach we use here in the neural network), the redondancy in measurement channels does not immediately translate into a higher measurement precision. However, as we shall see in the following, it does improve very strongly the resilience of the system to noise present in one or more channels.

\begin{figure}
	\centering
	\includegraphics[width=0.7\textwidth]{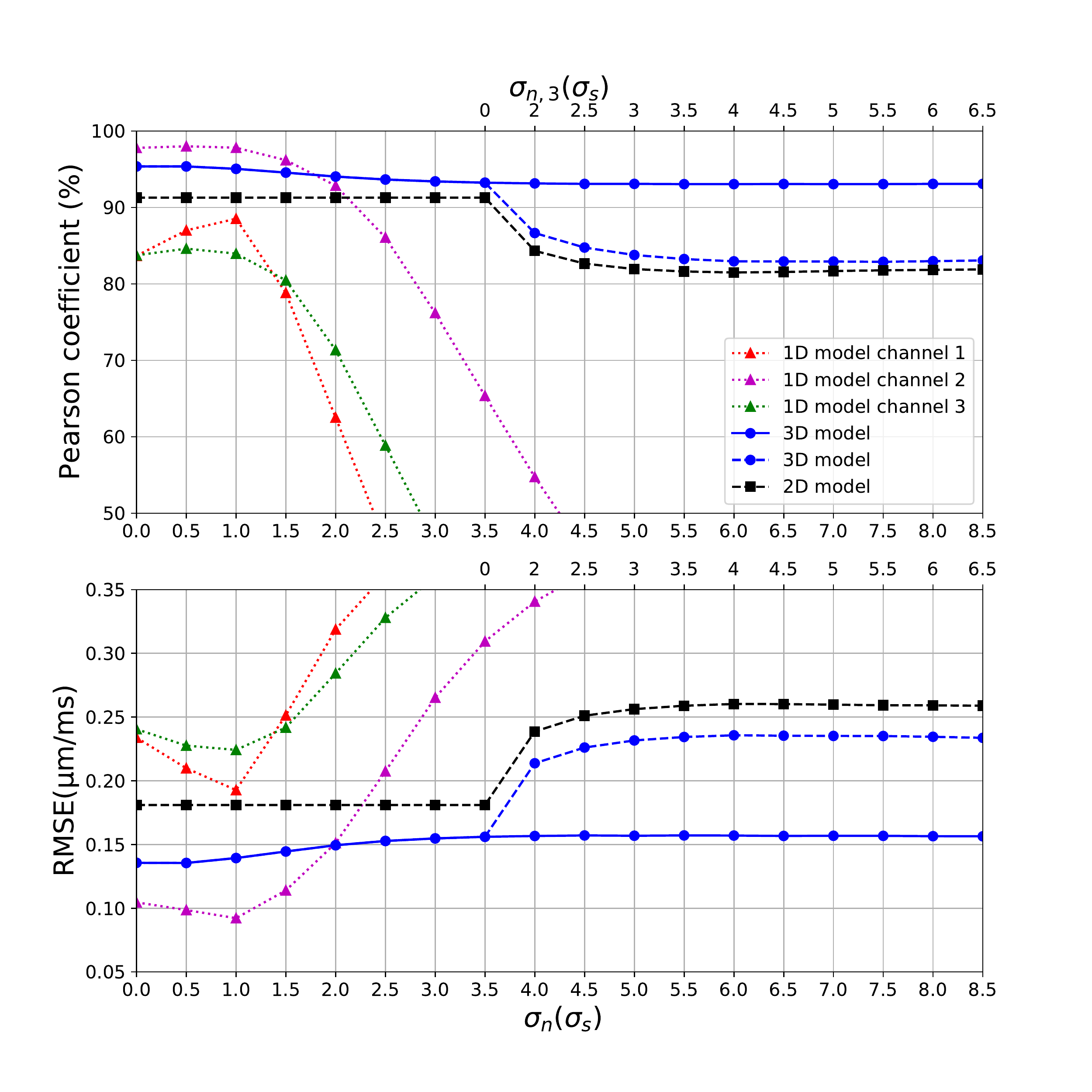}
	\caption{Robustness of the reconstruction against noise in one or more self-mixing signals. Top: Pearson correlation coefficient between the reconstruction provided by models and the true displacement. Bottom: Root Mean Squared Error on the displacement reconstruction. 
	\label{fig:noise}}
\end{figure}

To quantify this effect, we check the reconstruction performance on the same data set as above after adding to one or more channel(s) a $\delta$-correlated gaussian noise with standard deviation $\sigma_n$ which models the noise-induced degradation of self-mixing measurements. The results of this procedure are shown on \ffig{fig:noise}, where we show (top row) the correlation coefficient between the true displacement and the reconstruction of the displacement provided by several models and (bottom row) the Root Mean Square Error on the reconstruction for each model depending on noise added to one or more channels.

The three 1D models curves (red, green and purple triangles) are shown for reference and display essentially the same behavior. For each case, we use a single channel for the reconstruction and degrade this channel by adding noise to it. At first, each model provides an acceptable level of performance but even the best signal (channel 2) becomes very unreliable when the added noise is larger than $\sigma_n=3*\sigma_s$ where $\sigma_s$ is the standard deviation of the uncontaminated original self-mixing signal. The fact that there is an optimal non-zero level of noise in these curves is a result of our training procedure: for robustness all models have been trained with $\sigma_n=1$ and therefore have never seen uncontaminated signals.

The two curves based on 3D models (blue disks) show the robustness of a model operating on three channels simultaneously. The continuous blue line is obtained when noise is added only to the  second self-mixing channel \textit{ie} the one with the best performance when used in isolation. We observe that when a single channel is fully degraded ($\sigma_n=8\sigma_s$ for instance) this model still operates at a very high level of performance ($r=0.93, RMSE=0.16$). This is an excellent point in terms of measurement quality and availability. About measurement availability, it shows that a 3D measurement system can very well operate under highly degraded conditions \textit{ie} even when the highest performance channel is basically lost.
The dashed blue line is obtained when noise is added to the second self-mixing channel and also (starting from $\sigma_n>3.5\sigma_s$ to the third measurement channel. When this noise term $\sigma_{n,3}$ is added to the third channel, the measurement quality of course decreases since the model in the end operates basically on only a single remaining channel. Accordingly, when both channels 2 and 3 have become unusable due to very high noise (for instance $\sigma_n=7, \sigma_{n,3}=5$), the performance level of the 3D model is basically the performance level of the 1D model operating on the uncontaminated channel 1 $r=0.83, RMSE=0.23$. Again, this is an excellent point in terms of measurement availability since a three-channels measurement system can operate at the performance level of a single channel when the other two are lost.

Finally, for completeness, the performance of a 2D model (trained only on the poorest self-mixing signals 1 and 3) is shown as black squared dashed line. Of course the performance of that model is not affected by noise added to channel 2 which it does not process but it is important to underline that the 2D model performance is better than that of the 1D models based only on channels 1 and 3 in isolation. When noise is added also to channel 3, the performance of this model smoothly degrades down to that of the 1D model based only on an unperturbed channel 1.

Overall, the above observations demonstrate several outstanding properties (in terms of accuracy and robustness) of a 3D approach to displacement measurement:
\begin{itemize}
	\item In absence of noise, the 3D model's performance almost matches that of the best 1D model and is much better than the other two 1D models
	\item When noise degrades the most informative channel, the 3D model strongly outperforms all 1D models and also outperforms a 2D model based on unperturbed channels
	\item Even if noise strongly degrades two of the three available channels, the performance of the 3D model never degrades below that the 1D model based on the only unperturbed channel
	\item In presence of noise on all 3 channels, we checked that the 3D model performance is essentially equivalent to that of the best 1D model (channel 2), as was also observed in absence of noise (\ffig{fig:models}).
\end{itemize}

\subsection{Measurement availability}
\label{subsec:availability}

\begin{figure}
	\centering
	\includegraphics[width=0.7\textwidth]{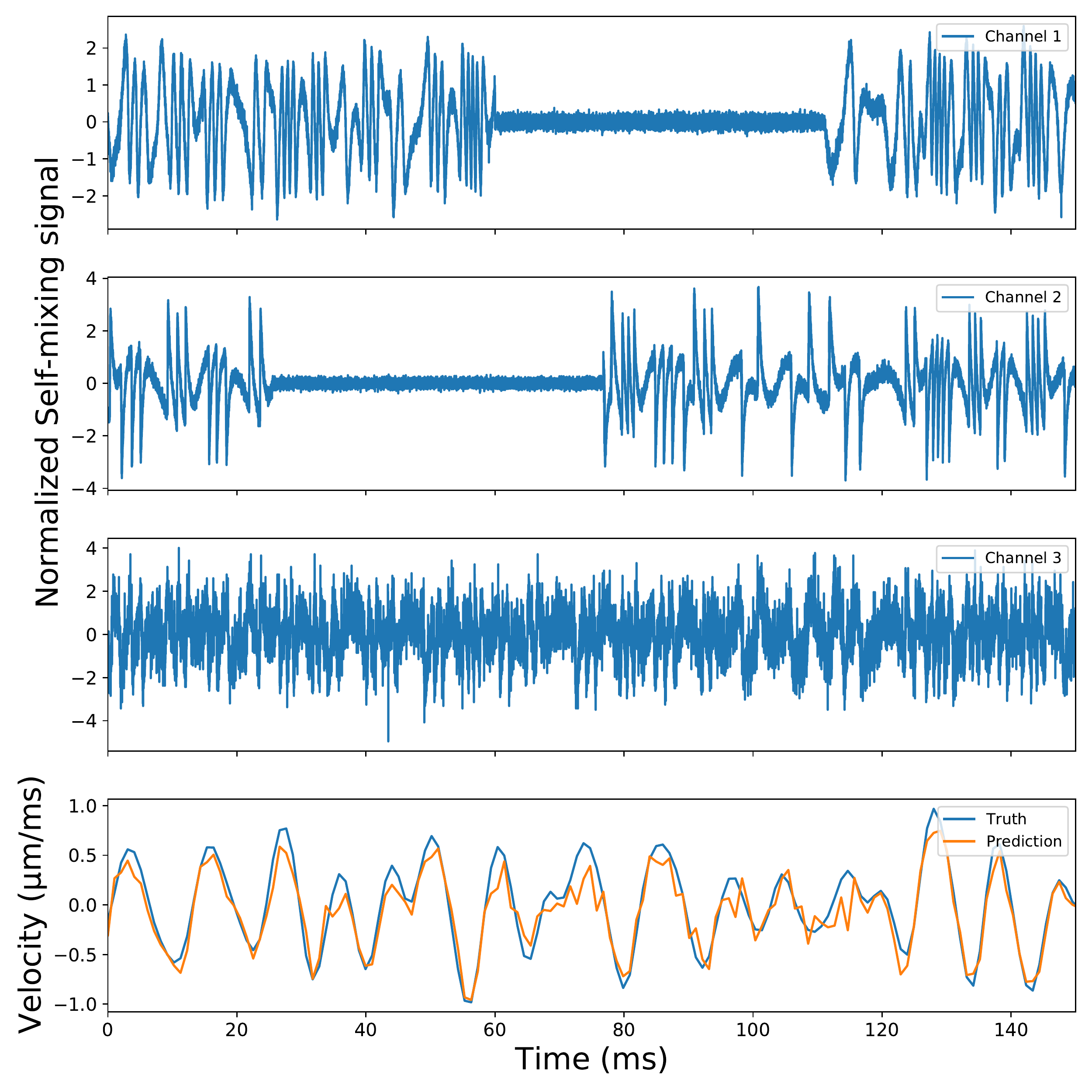}
	\caption{Channel loss and measurement availability. At different times, one or more self-mixing signals (top three rows) are replaced by white noise, simulating $C=0$ as could be caused for instance by speckle. The 3D model transparently makes use of the available data and provides (bottom row) a meaningful reconstruction of the displacement even in the worst case scenario of two channels becoming unavailable simultaneously. 
	\label{fig:availability}}
\end{figure}

The features described above have a considerable impact in terms of measurement availability, as we discuss below. Our goal here is to demonstrate that a displacement measurement system based on three simultaneous measurement channels processed by an adequately trained neural network provides a considerable step towards a high-availability self mixing displacement sensor. In fact, it is well known that with uncooperative targets, speckle can strongly modify the self-mixing signal shape, including leading to an effectively vanishing feedback rate. Several possibly complementary ways to mitigate this issue exist, including dedicated tracking hardware \cite{bernal2014robust}, dedicated algorithms \cite{siddiqui2017all}, on-the-fly parameter estimation \cite{bernal2021toward} or training a neural network in several feedback conditions to enable self-mixing signal processing in many operation ranges \cite{barland2021convolutional}. Here we focus on the case in which speckle (or in fact any other unwanted perturbation) leads to the full degradation of a self-mixing measurement channel.

To simulate this phenomenon, we replace one or the other channel by a gaussian white noise during a certain time interval (simulating the case of $C=0$ due to speckle or any other disturbance for this channel) and feed that modified interferometric data to the 3D model. The top three rows of \ffig{fig:availability} show the three self-mixing signals and the bottom row shows the true displacement and the reconstruction. This analysis is performed on the same segment of displacement as in \ffig{fig:example}. During the interval  $60<t<110$~ms, channel 1 is degraded and during the interval $25<t<75$~ms channel 2 is degraded, again simulating $C=0$. As we see, the 3D model always provides a meaningful reconstruction, even in the worst conditions such as the central region, during the interval $60<t<75$~ms where two channels out of three are unusable. Most importantly for a high-availability measurement system, the 3D model transparently makes optimal use of the available information but also the optimal quality of the reconstruction is recovered as soon as the quality of the self-mixing data itself is restored.

\section{Conclusion}
\label{sec:conclusion}

In conclusion, we have analyzed the performance of a high-availability displacement sensor based on three independent self-mixing interferometry channels processed by a lightweight neural network. Thanks to the inherent capacity of neural networks to process high-dimensional input, the multichannel sensor can transparently make optimal use of the self-mixing signal, using one or more input channels depending on their availability. The multichannel system performance nearly matches that of the best quality channel in absence of any disruption and as soon as one channel is degraded the multichannel sensor outperforms any single-channel sensor. Even when two channels are entirely degraded, the multichannel sensor still provides reliable displacement inference based on the only remaining channel. The network is trained to infer a displacement without relying on physical modelling, therefore the approach does not require any parameter estimation. Together with the ability of neural networks to learn many shapes of self-mixing signals \cite{barland2021convolutional}, we believe that the multi-channel approach constitutes a key element in solving the long-standing issue of speckle-affected self-mixing interferometry with non-cooperative targets.

Since the neural network is purposefully designed to process self-mixing data with near-minimal number of parameters, the network is amenable to embedding on tiny computing devices \cite{novac2021quantization,saha2022machine}, which opens the way towards small-footprint and low-power smart sensors leveraging the intrinsic simplicity of self-mixing interferometry. 

For future work, we underline that the approach outlined here is only meant as a proof of concept. First about the neural network itself: as in \cite{barland2021convolutional}, the network is extremely basic and can certainly be improved, especially through more advanced channel fusion. Also about the photonic stages, the present work opens perspectives along the lines of more advanced and multimodal sensing \cite{columbo2012self,brambilla2020versatile}, perhaps including compressed sensing \cite{li2022absolute} for onboarding on low-footprint components.

\begin{table}
	\centering
\begin{tabular}{ |p{2.5cm}|p{2.5cm}|p{2.5cm}|p{2.5cm}| p{2.5cm}| }
 \hline
 \multicolumn{5}{|l|}{Network structure: sequential} \\
 \hline
 Layer type & Main\newline hyperparameters &  Trainable\newline Parameters & Output shape & Activation function \\
 \hline
 1D convolutional   &  kernel size: 7\newline filters: 32    & 704&   (250, 32)& relu\\
 Max Pooling &  pool size: 2   & 0   &(125, 32)&\\
 1D convolutional & kernel size:7\newline filters: 64 & 14400&  (119, 64)& relu\\
 Max Pooling    & pool size: 2 & 0 &  (59, 64)&\\
 1D convolutional&   kernel size: 7\newline filters 128  & 57472 &(53, 128)& relu\\
 Max Pooling    & pool size: 2 & 0 &  (26, 128)&\\
 Dropout & 10\%  & 0   & (26, 128)&\\
 1D convolutional & kernel size: 7\newline filters: 128  & 114816 &(20,  128)& relu\\
 Max Pooling    & pool size: 2 & 0 &  (10, 128)&\\
 Flatten & & 0 &1280  & \\
 Fully connected & units: 16 & 20495 & 16 & relu\\
 Fully connected & units: 1 & 17 & 1 & linear\\
 \hline
\end{tabular}
	\caption{Main parameters of the network used in this work. The network is a sequence of 1-dimensional convolutional and dropout layers followed by two fully connected layers for the final regression. The total number of trainable parameters is 207 905.
		\label{tab:network}}
\end{table}

\section*{Appendix}
\label{sec:appendix}
The structure of the network is identical to that of \cite{barland2021convolutional} albeit with a larger number of convolutional kernels so as to accommodate three measurement channels instead of one. The key elements of the network are shown on Table 1. The network is implemented thanks to the Keras library \cite{chollet2015keras} and we refer the reader to deep learning fundamentals \cite{goodfellow2016deep} and implementations \cite{chollet2015keras} for background information. The total number of trainable parameters is 207 905. Networks of identical architecture with more cells per layer did not lead to significant improvements. The training of the network takes about twenty minutes on a consumer-grade GPU (NVIDIA GeForce GTX 1080).

\section*{Disclosures}

The authors declare no conflicts of interest.

\bibliography{selfmixing}

\end{document}